# Growth and Characterization of Bulk Aluminum Nitride by Physical Vapor Transport

Tim K. Hossain[1,2], James V. Lindesay[1,2], and Michael G. Spencer[2]

[1]Computational Physics Laboratory and [2]Materials Science Research Center of Excellence

Howard University, Washington, D.C.   20059

**ABSTRACT**

A major issue in the development of the technology of nitride based materials is the choice of substrate.  The structural and optical properties of the layers are intimately connected to the substrate material used in the epitaxial growth.  Key issues include lattice matching of substrate and epitaxial layer and the difference in the thermal expansion between the substrate and epitaxial layer.  Due to the extremely high vapor pressure of nitrogen over Gallium Nitride (GaN), pure GaN substrate are very difficult to produce and currently almost all nitride-based device are fabricated by hetero-epitaxy on either Sapphire or Silicon Carbide (SiC).  Alternatively, AlN has a significantly lower vapor pressure of nitrogen and can be grown in bulk form. In this paper we report on our results in the growth of AlN by physical vapor transport.  A theoretical model was developed to investigate the growth process by to determine optimal growth parameters.  Several different seed crystals were investigated, singular 6H-SiC, $3.5^0$ off-axis 6H-SiC and $8^0$ off-axis 4H-SiC, and AlN grown by hydride vapor deposition on silicon substrates (the silicon substrates were subsequently removed by etching). Auger electron microscopy, transmission electron microscopy, and various X-ray diffraction techniques were used to investigate the samples grown.

# **CHAPTER I:** Introductory Remarks

## **1.1 Introduction:**

The III-V nitride materials and their ternary alloys have great potential for use in optoelectronics and high-temperature electronic devices due to their wide range of band-gaps and high-temperature stability. These materials include gallium nitride (GaN), with a band-gap of 3.4eV, aluminum nitride (AlN), with a band-gap of 6.2eV, indium nitride (InN), with a band-gap of 1.9eV, and their alloys. Recently, these semiconductors have received much attention due to successful demonstration of high efficiency blue light-emitting diodes (LEDs).[1-3]

The blue and UV wavelengths are technologically important regions of the electromagnetic spectrum for which efforts to develop semiconductor device technology has been thus far unsuccessful. At present, semiconductor optical devices routinely operate from infra red (IR) to green wavelengths. If this region could be extended into blue wavelengths, semiconductor components could then emit and detect the three primary colors of the visible spectrum which would have a major impact on imaging and graphics application. However, the lack of suitable substrate materials that are both lattice and thermally matched to these nitride presents a serious obstacle to ongoing research and development efforts. In addition, it is well known that the crystal structure of deposited film is strongly dependent on the particular substrate materials.

The substrate materials of choice for most researchers are sapphire ($Al_2O_3$) and silicon carbide (SiC) with a large lattice mismatch with the nitride-based films. Minimizing the lattice mismatch and enhancing the growth conditions improve the nitride film morphology by lowering the density of defects such as misfit dislocations. Other defects, such as inversion domain boundaries, result from the film non-isomorphism with the substrate rather than the lattice mismatch.[4] The ideal substrate needs to be as lattice matched and isomorphic to the film as possible. Nitride buffer layers are often grown on sapphire or SiC in an attempt to provide a lattice matched and isomorphic surface for film growth. While buffer layers have helped improve film quality, film grown on buffer layers are usually columnar and contain a higher density of dislocations than desired. Due to the extremely high vapor pressure of nitrogen over gallium nitride (GaN), pure GaN substrate is very difficult to grow.[5] Although the exact nitrogen dissociation pressure is not known, values cited by Landolt and Bornstein[6] denote that the nitrogen dissociation pressure of AlN is orders of magnitude smaller than that of GaN or InN.[7] Bulk AlN, therefore, should be easier to grow than bulk GaN or InN.

**1.2 Literature Review:**

The first report of advances in relation to the growth of high-temperature single crystal AlN came from D. Hamilton in 1958 and K. Taylor and C. Leine in 1960.[8, 9] These researchers grew AlN by sublimation technique producing platelets 2-3mm across. The AlN platelets were grown by the vaporization of aluminum in nitrogen environment at $1800^0$C - $2000^0$C.[10] The success of Hamilton, Taylor, and Lenie growing AlN platelet led many researchers in the 1960's to follow the same techniques to study AlN.[11, 12]

The most satisfactory methods of growing high purity AlN single crystals appear to be those where AlN itself is used as the source material. Since high temperatures ($1500^0$C or above) are required, contamination from the crucibles is always a problem. Graphite crucibles or furnace tubes containing AlN have been used by many people for growing AlN crystal from the powder [9, [13, 14]. Extensive experiments on sublimation growth of AlN have been carried out by Champbell and Chang[14] and by Knippenberg and Verspui.[14, 15] In the Champbell and Chang experiments rather impure AlN powder containing carbon impurity was used, temperatures of $1950^0C$ to $2200^0C$ were employed and graphite crucibles were used. Plate-like crystals up to 1mm across were produced. The experiments of Knippenberg and Verspui were carried out between $1800^0C$ and $2300^0C$ in graphite crucibles. A crystal up to few a millimeter in size was produced.

In 1976, Slack and McNelly[16] had achieved the largest crystal (10mm long by 3mm diameter) using the sublimation-recondensation method at $2250^0C$. Prior to the present research, C.M. Balkas and R.F. Davis[17], had grown single crystals of AlN to 1mm thickness in the range of $1950^0C$ - $2250^0C$ at 500 Torr of $N_2$ on 10x10 $mm^2$ 6HSiC (0001) substrate via sublimation-recondensation method. Close source to-substrate distance and an $80^0C$ - $150^0C$ temperature gradient were necessary to achieve this result.

## 1.3 Motivation for Present Work:

The present research is the next step toward achieving a better quality bulk AlN. The study describes the growth bulk aluminum nitride (AlN) on 6HSiC substrate both off-axis and on-axis by physical vapor transport. The growth of AlN is investigated in the temperature range of $1900^0C$ - $2200^0C$ with nitrogen pressure of 200 - 500 Torr. Growth material is characterized by several standard techniques such as, auger electron spectroscopy (AES), X-ray diffraction, secondary electron microscopy (SEM) and high-resolution transmission electron microscopy (TEM).

In the second part of paper, a mathematical model have been developed to understand the detailed growth process of our system. The variation of growth rates with temperature and pressures have been examined for our particular system. In the end, the analytical result is compared with the experimental result.

## CHAPTER II: Experimental Details

### 2.1 Introduction:

Bulk AlN crystal growth was accomplished using the Howard University Sublimation Reactor. Crystal growth was performed using a close spaced sublimation growth technique.[18] In this method the source and the substrate is separated by a distance 1-4 mm and the growth mechanism is physical vapor transport (PVT). A systematic crystal growth scheme has been developed in order to successfully grow reproducible bulk AlN single crystals. The essential experimental details are provided in the following sections.

### 2.2 Sublimation System:

In this section, the experimental set up used in the physical vapor transport (PVT) growth of AlN is illustrated. The main reactor is a water cooled stainless steel cylindrical chamber capable of operation under high vacuum levels. Within the reactor is a cylindrical thermal insulation shield 59.94indiameter and 76.2cm long made of graphite. The shield has a 10.16cm diameter and 43.18cm long central cavity capable of housing the heater elements and the growth crucible. The heating arrangement itself consists of two resistively heated high purity graphite spiral heating elements each 17.78cm long placed one on top of the other. The upper filament is held in position with the help of two detachable graphite arms. Likewise, the lower element is supported on two detachable graphite legs which is referred to as zone 2. The hollow space within the spiral heating element is used to contain the growth crucible. The crucible attached to the crucible support rod is conveniently inserted into the growth chamber via a port located at the top of the chamber. Figure 2.1 shows the internal reactor configuration.

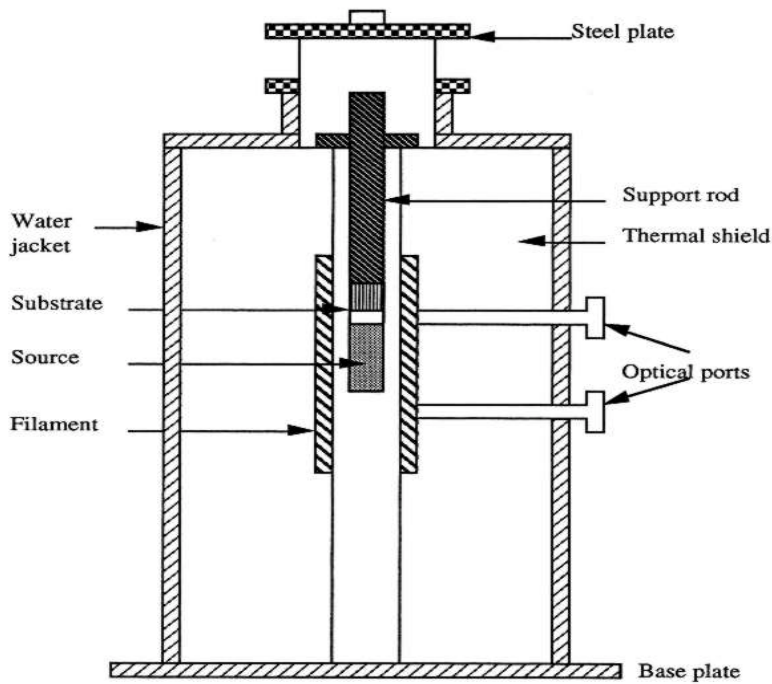

Figure 2.1 Sublimation reactor

The reactor is designed to operate up to $2400^0C$ and a temperature controller maintains the set point temperature during the growth period using two pyrometers. Holes bored through the thermal shield allow the pyrometers to directly view the internal cavity at the midpoint of each heater element. Crystal growth is carried out in a nitrogen ambient. The nitrogen pressure during various stages of growth was monitored using a combination of a penning, cold cathode discharge gauge, with a $10^{-2}$ - $10^{-7}$ mbar range and a Barton transducer with a $10^{-1}$ - $10^{-3}$ torr range. A throttle value interlocked in a Barton transducer feedback loop is employed to regulate the chamber pressure during growth. A diffusion pump connected to the main chamber helped achieve high level of vacuum (low $10^{-5}$ torr). All system operations were performed manually during the growth process.

**2.3 Crucible Configuration**:

The most critical part of the physical vapor process is the crucible structure. All crucible parts were fabricated from ultra high purity graphite and the interior sides were coated with SIC (2000°A) by a commercial vender. The crucible has a cylindrical symmetry with an outer diameter of 2.54cm, and has two separable parts. The bottom part, used to contain the source powder, is referred to as the crucible bottom (CB). The top portion which can be attached to the CB by means of screw threads, (called the transfer tube) is designed to hold the substrate holder (the pedestal). The substrate (6HSiC) mounted on the pedestal top. Thus, the substrate is placed near the top of the source powder at a separation 2-3 mm. The entire assembly constitutes the crucible with a total height of 6.35 cm (2.5 inches). Each pedestal has 4 side "ribs" in order to partially open the growth cell. The height of CB is 3.81 cm (1.5 inches) which is also the source powder height. A temperature gradient is imposed across the growth crucible with the aid of the graphite support rod which acts to transport thermal energy to a heat sink at the top of the furnace. An understanding of which elements might act as impurities in AIN is important in growing crystals of high purity. It is found that both carbon (C) and oxygen (O) is acting as highly soluble impurities in AIN by substituting for N. This problem has been addressed by modifying the crucible bottom (CB) part, and the source material. It is assumed that the source of carbon contamination, is due to the CB itself which is made of graphite. Thus, the CB walls were coated with a 1-5 micron commercial tungsten foil prior to filling it with AlN powder. Oxygen contamination is reduced by reusing the same AlN powder. It was observed that the AlN powder was molded and became AlN powder chalk after a growth. Since the surface area of AlN chalk is less than the surface area of AlN powder, a smaller area is exposed to react with the moisture and/or oxygen in the air to form a surface layer of aluminum oxide. Therefore, AlN chalk reduced the carbon contamination and was a better source material. In our studies, crystal grown through cycle 34 onwards showed these changes. A dramatic improvement in crystal quality was achieved during this phase of the research.

## 2.4 Furnace Temperature Profile:

A furnace temperature profile was performed to determine as accurately as possible the axial temperature distribution along the column where the growth crucible is positioned, using a thermocouple (T/C) enclosed in a molybdenum sheath. Plots showing temperature versus distance along crucible axes helped determine the temperature gradient between the source top and the substrate. The plot was interpreted in conjunction with the external pyrometer set point readings. This was facilitated by using a dummy graphite crucible with a central cavity to contain the thermocouple. Either zone 1 (Z1) or zone 2 (Z2) was employed depending on need. Thermocouple readings were noted at spots where the actual crucible would rest. Measurements were performed twice at each spot, once during the T/C extraction and back during its insertion. The average reading was used as the data point for plotting. Measurements were performed at 1.27 cm (half inch) interval until the entire crucible range was covered.. Measurements were performed for controller set point temperature of 1900ºc, 2000ºc and 2100ºc respectively. System pressure was maintained at 500 torr ($N_2$ ambient) during the profile. Most of the growth experiments were performed at this pressure. The profiles helped in understanding the thermal behavior of the system for a particular external set temperature. Temperature gradients determined from these curves were used to locate substrate position in the cavity. Both zones (Z1 and Z2) were employed during crystal growth cycles. Measurements were made with CB position coinciding with zone 1 center. The substrate position is 5.08 cm(2 inches) above the CB Hence for a particular set point temperature, seed temperature can be estimated form the temperature curve as shown in figure 2.2. A tangent drawn at the point on the curve help estimates the temperature gradient. The temperature gradient was estimated at 2~3ºC per mm for a substrate-source separation of 3~4 mm. As a result of failure of the Z1 filament, growth experiments were performed using only z2 filaments and the corresponding thermal profile is shown in Fig. 2.2. The z2 center is located 12.7 cm (5 inches) below z1 center. The set point temperatures were maintained at 2000ºC, 2100ºC and 2205ºC as before. A temperature gradient in the range of 1.5ºC to 1.75ºC per mm was achieved for a 3 mm seed-source separator. This corresponds to a temperature difference range of 4.5 to 5.25. Therefore, there is an uncertainty in the temperature difference measurement.

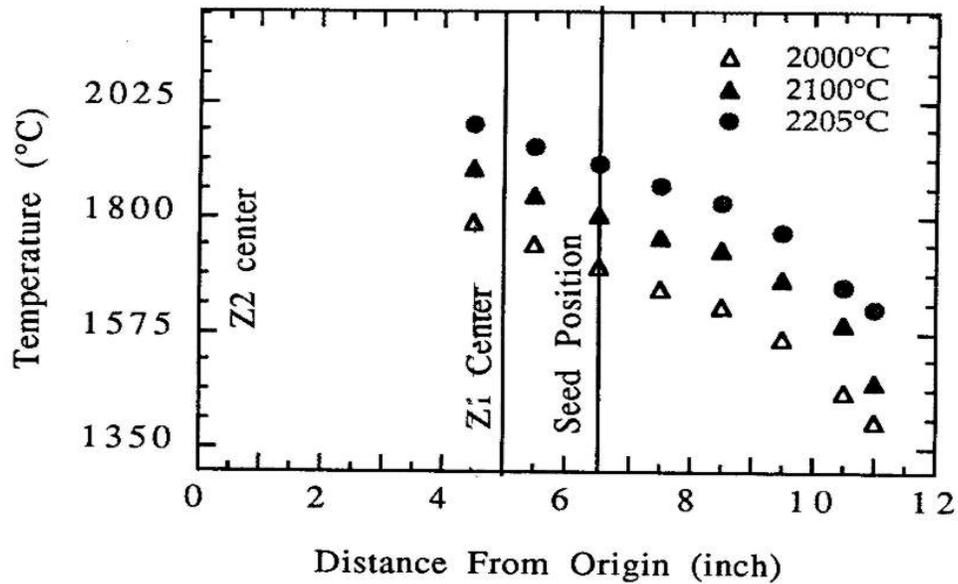

Figure 2.2 Temperature profile 2

## 2.5 Physical vapor Crystal Growth:

Crystal growth was performed using the Howard University sublimation reactor. Growth was performed in the substrate temperature range of 1900°C to 2230°C during various stages of research. Nitrogen ($N_2$) gas was used as the ambient during the process. System pressure was varied from 200 to 600 torr. Once the crucible is placed within the chamber, it is allowed to attain initial vacuum levels of $10^{-2}$ torr using a mechanical pump. This is followed by placing the chamber on the diffusion pump (lower $10^{-5}$ torr) and performing the initial bake out for one hour at 1200°C by setting zone current controller switch at the 50 amps position. At the completion of the bake out, the diffusion pump value is closed and the chamber by is pressurized by $N_2$ gas to 500 torr. A ten minute waiting period for pressure stabilization is followed by increasing the current setting to maximum level. This corresponds to a zone temperature of 1370°C. The controller set point temperature is then increased to 1900ºC. At this point, mechanical pump value is opened to maintain the chamber pressure at the desired level. A 100ºC increment is made every 5 minutes to the desired growth temperature (usually 2100ºC ~ 2200⁰C ). The actual growth period is counted from this point on and maintained for one hour. After the growth cycle is completed, furnace power is switched off and the chamber is cooled under vacuum.

# CHAPTER III: Theoretical development

## 3.1 Introduction:

As mentioned earlier, our system of crystal growth was a closed spaced growth technique, in which the source and the substrate are separated by a small distance and the growth mechanism is physical vapor transport. $N_2$ gas was used as the ambient gas during the process. The growth was performed in a temperature-gradient zone, so that there appears a certain temperature difference $\Delta T$ between the source (AlN) and the substrate (6HSiC).

## 3.2 Outline of the Model:

The mathematical model is developed for a one-dimensional system, in which a crystal surface is growing at x=l, and the component vapor are introduced at x=0. The model assumed that the following parameters can be controlled:

1. the total pressure $P_T$ of the system and
2. the source temperature $T_S$.

In our model a crystal of an $A^{III}B^V$ compound in equilibrium with its vapor of A atoms and $B_m$ molecules is considered. Let the equilibrium partial pressure of A atoms and $B_m$ be $P_A^*$ and $P_{Bm}^*$, which are functions of absolute temperature T. The dissociative sublimation of the solid $AB_m$ can be written according to the equation.[26]

$$AB_n \longrightarrow A(g) + \left(\frac{n}{m}\right) B_m(g) \tag{1}$$

Also at equilibrium, it can be written[27, 28].

$$\frac{P_A^* P_{B_m}^{*\frac{1}{m}}}{a_{AB_n}} = K_T = e^{\left(-\frac{\Delta H}{RT} + \frac{\Delta S}{R}\right)} \tag{2}$$

where $a_{AB}$ is the activity of the solid $AB_n$, $K_T$ is the equilibrium constant, and $\Delta H$ and $\Delta S$ and are the molar enthalpy and entropy of sublimation. The equilibrium is generally of a dynamic nature, representing a balance of condensation and sublimation rates. Condensation and sublimation rate can be described by Langmuir adsorption and desorption in terms of the activation energy Q for adsorption, the enthalpy of adsorption E, and the coverage $\theta$ on the surface. The detail calculation of the Langmuir absorption and desorption are presented in the following section.

### 3.3 Adsorption, physisorption and chemisorption:

Atoms at the surface of a solid exert an attractive force normal to the surface plane.

Consequently, at a gas/solid interface, the concentration of gas exceeds that in the gas phase. The excess concentration at the surface is termed adsorption; the solid is the adsorbent and the gas the adsorbate. The amount adsorption depends on the physical and chemical properties of both adsorbent and adsorbate, the temperature of the gas/solid system and the ambient pressure of the gaseous adsorbate. There are two kinds of adsorption, physisorption and chemisorption. In physisorption, the adsorbate is held to the surface by the Van der Waals forces that arise primarily from the interaction of fluctuating dipoles. Chemisorption is considered to be a chemical reaction between an adsorbate molecules and the surface atom (adsorbent molecules).

In physisorption, the interaction can be described by the potential-energy diagram shown in figure 3.1. An incoming molecule with kinetic energy $E_K$ has to lose at least this amount of energy in order to stay on the surface. It loses energy by exciting lattice phonons in the substrate and the molecule then comes to equilibrium in a state of oscillation in the potential well of depth equal to the binding or adsorption energy $E_A$. In order to leave the surface the molecule must acquire enough energy to surmount the potential-energy barrier $E_A$ (activation energy).

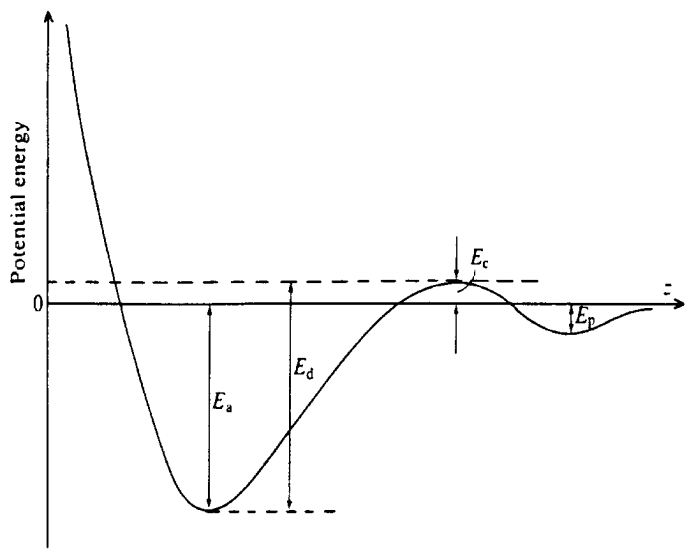

Figure 3.1 Potential energy diagram for chemisorption on planar surface

The chemisorption occurs when integral numbers of electrons leave the adsorbate molecule and stay on the nearest substrate atom or vice versa. There is an admixture of the wave functions of the valance electrons of the adsorbate molecule with the valance electrons of the substrate (adsorbent) into a new wave function. The electrons responsible for the bonding can then be thought of as moving in orbitals between substrate and adatoms (adsorbate molecule atom) and a covalent bond is formed. The potential energy diagram for chemisorption is shown in figure 3.1. Some of the impinging adsorbate molecules are accommodated by the surface and become weakly bound in a physisorbed sate (called precursor state) with binding energy $E_P$. During their stay time in this state vibrational processes can occur which allow then to surmount the energy barrier ($E_C + E_P$) and electron exchange occurs between the adsorbate and substrate (adsorbent) where each adatom now finds itself in a much deeper well $E_a$, as shown in figure 3.1. It is chemisorbed.

## 3.4 Theoretical Model:

In moving from the gas phase to the adsorbed layer the adsorbate molecules passes over a potential energy barrier, the height of which is the energy of activation. In the transition state, the activated complexes vibrate along the coordinate perpendicular to the surface with frequency $v$. This mode of vibration has the feature that as the molecule moves either away from or towards the surface, its energetics affects the stability of the state. The frequence $v$ is therefore the frequency of decomposition of the complexes and if C* is their concentration per m² of surface, the velocity of adsorption per unit area is

$$u = vc^* \tag{3}$$

Similarly the velocity of desorption well be

$$u' = vc^{**} \tag{4}$$

If $N_g$ is the number of molecules in the gas phase of volume V m³, and $N_s$ and N* are the number of bare sites and transition complexes on a surface of area S per m², the following concentrations can be defined:

$$C_g = \frac{N_g}{V}$$

$$C_s = \frac{N_s}{S}$$

$$C^* = \frac{N^*}{S}$$

where $C_g$ is gas molecules/m³, $C_s$ is bare sites/m², and $C^*$ is transition complexes/m². The equilibrium constant on a uniform surface between complexes, gas molecules and bare sites can be written as

$$k^* = \frac{C^*}{C_g C_s} = \frac{N^*}{\frac{N_g}{V} N_s} \tag{5}$$

The ratio of concentration ( $C_g$, $C_s$, $C^*$ ) is equal to the ratio of probabilities of the molecules existing as gas molecules, bare sites or transition complexes, also related to the ratio of partition functions. Equation (5) can be expressed as

$$\frac{C^*}{C_g C_s} = \frac{N^*}{\left(\frac{N_g}{V}\right) N_s} = \frac{f^*}{\left(\frac{f_g}{V}\right) f_s} \tag{6}$$

where the f's are the partition functions for the species. Representing the partition function of the gas per unit volume ($f_g/V$) as $F_g$, and separating the zero point energies from the partition functions, equation(6) becomes

$$\frac{C^*}{C_g C_s} = \frac{f^*}{F_g f_s} e^{-\frac{Q}{RT}} \tag{7}$$

where Q is the activation energy of the adsorption.

In the function $f^*$, there is a term due to the vibration $f_v$, perpendicular to the surface. This frequency of the vibration is expected to be small, which allows the function $f_v$ to be written in the approximate form $f_v \approx \frac{kT}{h\upsilon}$.

Substituting this value, equation(7) becomes

$$\frac{C^*}{C_g C_s} = \frac{kT}{h\upsilon} \frac{f^*}{F_g f_s} e^{-\frac{Q}{RT}} \qquad (8)$$

The velocity of chemisorption is then found from equation(3).[29]

$$u = C_g C_s \frac{kT}{h} \frac{f^*}{F_g f_s} e^{-\frac{Q}{RT}} \qquad (9)$$

For chemisorption, the molecules translate freely only within the area of a site, and undergo activation jumps to neighboring sites. In this case[29]

$$C_g kT = \frac{N_g}{V} kT = P$$

$$f_s = 1$$

$$C_s = N_s f(\theta)$$

$$F_g = \frac{(2\pi m kT)^{\frac{3}{2}}}{h^3} b_g$$

where $N_s$ is the number of surface sites/m², $f(\theta)$ is a fraction of surface site available for chemisorption and $b_g$ is the partition function due to rotation and vibration of the molecules.

The complex possesses two degrees of translational freedom within the area A m² occupied by a site where A = $1/N_s$, and also has the same rotational and vibrational freedom as in the gas phase. Therefore

$$f^* = \frac{2\pi m kT}{h^2} \frac{1}{N_s} b_g$$

Inserting these values in equation(9), we have

$$u = \frac{P}{\sqrt{2\pi m RT}} f(\theta) e^{-\frac{Q}{RT}} \qquad (10)$$

This is the velocity of adsorption which is referred to as Langmuir adsorption isotherm.

In deriving the velocity of desorption, statistical equilibrium is assumed between adsorbed molecules and activated complexes. Again, activated molecules vibrate perpendicular to the surface with a frequency $v$, which is the frequency of decomposition of the complexes. By a similar argument, the velocity of desorption depends on three factors; the number of sites, f'($\theta$), from which desorption is possible, the activation energy $E'$, and a relative flux factor K. The velocity of desorption, $u'$ is then

$$u' = Kf'(\theta)e^{-\frac{E'}{RT}} \qquad (11)$$

As shown in figure 3.2, $E'$ is equal to the sum of the activation energy Q, and heat of adsorption E

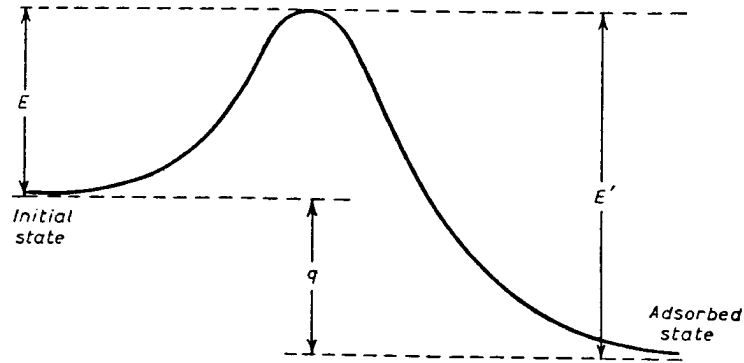

Figure 3.2 Activation energy diagram

Now using the Polanyi-Wigner equation[29] we can write

$$K = N_s v$$

where $N_s$ is the number of surface sites/m². Using this value, equation(11) becomes

$$u' = N_s v f'(\theta) e^{-\frac{(Q+E)}{RT}} \qquad (12)$$

This is the velocity of desorption known as the Langmuir desorption isotherm.

Representing the coverage by $\theta$, the fractional number of uncovered single sites is $(1-\theta)$. The chance that a colliding molecule strikes a vacant site for mobile and immobile layers is the same[29]

$$f(\theta) = 1 - \theta$$

Similarly, desorption can proceed from any single occupied site, and hence[26]

$$f'(\theta) = \theta$$

Substituting this values in equation(10) and equation(12) we obtain

$$u = \frac{P}{\sqrt{2\pi m RT}} (1-\theta) e^{-\frac{Q}{RT}}$$

$$u' = N_s \upsilon \theta e^{-\frac{(Q+E)}{RT}}$$

(13)

Considering the reaction equation(1), we can write both adsorption and desorption for component A as follows

$$u = \frac{P_A^*}{\sqrt{2\pi m_A RT}} e^{-\frac{Q_A}{RT}}$$

$$u' = N_A \upsilon_A \theta_A e^{-\frac{(Q_A + E_A)}{RT}}$$

(14)

and for component $(B_m)$

$$u = \frac{P_B^{\frac{1}{m}}(1-\theta_m)}{(2\pi m_{B_m} RT)^{\frac{1}{m}}}$$

(15)

$$u' = N_{B_m}^{\frac{1}{m}} \upsilon_{B_m}^{\frac{1}{m}} \theta e^{-\frac{(Q_B+E_B)}{RT}}$$

where $N_A$, $N_B$ are the density of surface sites for adsorption of component A and $B_m$ respectively. At equilibrium the rate of condensation is equal to the rate of desorption: $u = u'$. Thus using equation(14) and equation(15) it can be written

$$\frac{P_A^*(1-\theta_A)}{\sqrt{2\pi m_A RT}} e^{-\frac{Q_A}{RT}} = N_A \upsilon_A \theta_A e^{-\frac{(Q_A+E_A)}{RT}}$$

(16)

$$\frac{P_B^*(1-\theta_{B_m})}{(2\pi m_{B_m})^{\frac{1}{m}}} e^{-\frac{Q_{B_m}}{RT}} = N_{B_m}^{\frac{1}{m}} \upsilon_{B_m}^{\frac{1}{m}} \theta_{B_m} e^{-\frac{(Q_{B_m}+E_{B_m})}{RT}}$$

(17)

If a finite departure of $P_A$ and $P_B$ from their equilibrium values is now considered, the growth rate J can be equated to the imbalance of condensation (adsorption) and sublimation (desorption) rates for each component (A and $B_m$):

$$J_A = \frac{P_A(1-\theta_A)}{\sqrt{2\pi m_A RT}} e^{-\frac{Q_A}{RT}} - N_A \upsilon_A \theta_A e^{-\frac{(Q_A+E_A)}{RT}}$$

(18)

and for $B_m$

$$J_B = \frac{P_{B_m}^{\frac{1}{m}}(1-\theta_B)}{(2\pi m_{B_m} RT)^{\frac{1}{2m}}} e^{-\frac{Q_B}{RT}} - N_{B_m}^{\frac{1}{m}} v_{B_m}^{\frac{1}{m}} \theta_B^{\frac{1}{m}} e^{-\frac{(Q_B+E_B)}{RT}} \tag{19}$$

Using equation(16) and equation(17) we can solve for the coverage

$$\theta_A = \frac{P_A^*}{P_A^* + N_A v_A \sqrt{2\pi m_A RT} \, e^{-\frac{E_A}{RT}}} \tag{20}$$

Similarly,

$$\theta_B = \frac{P_{B_m}^{*\frac{1}{m}}}{P_{B_m}^{*\frac{1}{m}} + N_{B_m}^{\frac{1}{m}} v_{B_m}^{\frac{1}{m}} (2\pi m_{B_m} RT)^{\frac{1}{2m}} e^{-\frac{E_B}{RT}}} \tag{21}$$

Using equation(16) and equation(20), equation(18) can be written as

$$J_A = \frac{(P_A - P_A^*)}{\sqrt{2\pi m_A RT}} \left( \frac{N_A v_A \sqrt{2\pi m_A RT} \, e^{-\frac{E_A}{RT}}}{P_A^* + N_A v_A \sqrt{2\pi m_A RT} \, e^{-\frac{E_A}{RT}}} \right) e^{-\frac{Q_A}{RT}}$$

We will denote

$$k_A = N_A v_A \sqrt{2\pi m_A RT} \, e^{-\frac{E_A}{RT}} \tag{22}$$

Then growth rate $J_A$ becomes

$$J_A = \frac{(P_A - P_A^*)}{\sqrt{2\pi m_A RT}} \frac{k_A}{k_A + P_A^*} e^{-\frac{Q_A}{RT}} \tag{23}$$

Similarly for component $B_m$, growth rate will be

$$J_B = \frac{\left(P_{B_m}^{\frac{1}{m}} - P_{B_m}^{*\frac{1}{m}}\right)}{\left(2\pi m_{B_m} RT\right)^{\frac{1}{2m}}} \frac{k_B}{k_B + P_{B_m}^{*\frac{1}{m}}} e^{-\frac{Q_B}{RT}} \tag{24}$$

where

$$k_B = N_{B_m}^{\frac{1}{m}} \nu_{B_m}^{\frac{1}{m}} \theta_B \left(2\pi m_{B_m} RT\right)^{\frac{1}{2m}} e^{-\frac{E_B}{RT}} \tag{25}$$

The partial pressures of $P_A$ and $P_{B_m}$ at equilibrium can be calculated from equation(23) and equation(24) as follows:

$$P_A^* = \frac{P_A - J\sqrt{2\pi m_A RT}\, e^{\frac{Q_A}{RT}}}{1 + \frac{Je^{\frac{(Q_A + E_A)}{RT}}}{N_A \nu_A}} \tag{26}$$

and

$$P_{B_m}^* = \frac{P_{B_m}^{\frac{1}{m}} - J\left(2\pi m_{B_m} RT\right)^{\frac{1}{2m}} e^{\frac{Q_B}{RT}}}{1 + \frac{Je^{\frac{(Q_B + E_B)}{RT}}}{N_{B_m}^{\frac{1}{m}} \nu_{B_m}^{\frac{1}{m}}}} \tag{27}$$

where we have use the continuity condition $J = J_A = J_B$.

Finally, substituting equation(26) and equation(27) into equation(2), we obtain

$$\left[\frac{P_A - J\sqrt{2\pi m_A RT}\, e^{\frac{Q_A}{RT}}}{1 + \frac{Je^{\frac{(Q_A+E_A)}{RT}}}{N_A \nu_A}}\right]\left[\frac{P_{B_m}^{\frac{1}{m}} - J(2\pi m_{B_m} RT)^{\frac{1}{2m}} e^{\frac{Q_B}{RT}}}{1 + \frac{Je^{\frac{(Q_B+E_B)}{RT}}}{N_{B_m}^{\frac{1}{m}} \nu_{B_m}^{\frac{1}{m}}}}\right] = K_T \quad (28)$$

Equation(28) is the required equation, which expressed a relation between the growth rate, the actual partial pressures at the growing interface, and kinetic parameters.

The partial pressures $P_A$ and $P_{Bm}$ can be calculated from the theory of gas phase transport in the system. In our model, the component vapors move down the one-dimensional system from x=0 to x=l. Since the ends of the tube have different temperatures, there will be a pressure gradient between x= 0 to x= l, with a mass flow towards the low-pressure end. The mass flow of the component gas can be written in terms of a flow velocity v m/s, which is the drift velocity of the gas molecules due to the pressure difference.

$$J_1 = \frac{vP}{RT} \quad (29)$$

For a two component system, we can write equation(29) as

$$J_{1A} = \frac{vP_A}{RT}$$

$$J_{1B} = \frac{mvP_{B_m}}{RT} \quad (30)$$

where we have considered a general case with A monatomic and B m-atomic in the vapor phase.

Equation(30) is equivalent to saying that the gas mixture as a whole moves with a velocity v, and that the flux J of any component at any point is proportional to v and to the partial pressure of that component at the point under consideration.

If the ratio of the partial pressures $P_A/P_{Bm}$ at x=l is not the same as the stoichiometric composition of the solid, a build up of one component and a depletion of the other one will occur near x=l, which will lead to concentration gradients. Thus there will be a second contribution to the flux of each component caused by diffusion.

$$J_{2A} = -D_A \frac{dN_A}{dx}$$

(31)

$$J_{2B} = -mD_B \frac{dN_{B_m}}{dx}$$

where the D parameters are the diffusion coefficients of the components A and B. According to our system, the temperature gradient (dT/dx) is very small. From the equation of state for an ideal gas we have

$$\frac{d}{dx}\left(\frac{n}{V}\right) = \frac{1}{RT}\frac{dP}{dx} - \frac{P}{RT^2}\frac{dT}{dx}$$

where N = n/V is the density or concentration of the molecules. For small temperature gradients

$$\frac{dN}{dx} = \frac{1}{RT}\frac{dP}{dx} \tag{32}$$

Substituting equation(32) into equation(31) and adding equation(30), we can obtain the net flux of components A and B

$$J_A = \frac{vP_A}{RT} - \frac{D_A}{RT}\frac{dP_A}{dx} \tag{33}$$

$$J_B = \frac{mvP_{B_m}}{RT} - \frac{mD_B}{RT}\frac{dP_{B_m}}{dx} \tag{34}$$

The total pressure of our system is constant, thus we have

$$P_A + P_{B_m} = P_T \tag{35}$$

$$\frac{dP_A}{dx} = -\frac{dP_{B_m}}{dx} \tag{36}$$

where $P_T$ is a total constant pressure of the system. Since $P_T$ is constant, the drift velocity of the gas must be constant, or else the gas would accumulate where v was smaller and become subtle where v was larger. Moreover, $J_A$ and $J_B$ must be constant, because there is no source or sink of either component in our model, except for a sink at x = l and a source at x = 0. Now considering V, T, the J's and the D's all constant and integrating equation (33), we get

$$\ln\left(P_A - \frac{J_A RT}{v}\right) = \frac{vx}{D_A} + const \tag{37}$$

A similar equation can be obtained for component B. Applying boundary condition,

$$P_A = P_A^0$$

$$P_{B_m} = P_{B_m}^0$$

at x = 0, the constant of integration can be eliminated. Thus, at $P_A = P_A^0$, equation (37) becomes

$$\ln\left(P_A^0 - \frac{J_A RT}{v}\right) = const$$

Substituting this value in equation(37) becomes

$$P_A = P_A^0 e^{\frac{vx}{D_A}} - \frac{J_A RT}{v}\left(e^{\frac{vx}{D_A}} - 1\right) \tag{38}$$

Similarly for component $B_m$

$$P_{B_m} = P_{B_m}^0 e^{\frac{vx}{D_B}} - \frac{J_{Bm} RT}{mv}\left(e^{\frac{vx}{D_B}} - 1\right) \tag{39}$$

Adding equation(38) and equation(39) we have

$$P_A + P_{B_m} = \left(P_A^0 - \frac{J_A RT}{v}\right)e^{\frac{vx}{D_A}} + \left(P_{B_m}^0 - \frac{J_{Bm} RT}{v}\right)e^{\frac{vx}{D_B}} (J_A + J_{Bm})\frac{RT}{v} \tag{40}$$

The left-hand side is a constant, $P_T$. On the right-hand side, the last term is constant, but the first two terms are functions of the position x. In order for equation(40) to hold for all values of x, $D_A$ and $D_B$ must be equal. This fact is also known from the binary mixtures of components.[30]

$$D_A = D_B = D \tag{41}$$

Also

$$P_A + P_{B_m} = P_A^0 + P_{B_m}^0 \tag{42}$$

Then equation(40) becomes $P_T = \frac{(J_A + J_{Bm})RT}{v}$. The flux $J_{Bm}$ can be written as

$$J_{Bm} = \frac{1}{m} J_B. \text{ Then we have}$$

$$P_T = \frac{\left(J_A + \frac{J_B}{m}\right)RT}{v} \tag{43}$$

Let us now consider the equation(1) for the reaction of binary compound $AB_n$.

The condition for the components A and B to arrive at the substrate in the correct ratio is

$$J = J_A = J_{Bn} = \frac{1}{n} J_B \qquad (44)$$

Substituting equation(44) in equation(43) and rearranging, it is found that

$$v = \frac{JRT}{P_T}\left(1 + \frac{n}{m}\right) \qquad (45)$$

The factor $(1 + n/m)$ is a number that depends on the atomicity of B in the vapor, and the stoichiometric ratio in the solid. For aluminum nitride (AlN) grown from Al and $N_2$, n=1 and m=2, which gives $v = \frac{3JRT}{2P_T}$. Inserting this value of $v$ in equation(38) and equation(39), the partial pressures of components can be obtained:

$$P_A = \left[P_A^0 - \frac{2P_r}{3}\right] e^{\frac{3JRT}{2P_T D} + \frac{2P_T}{3}} \qquad (46)$$

$$P_{B_m} = \left(P_{B_m}^0 - \frac{P_T}{3}\right) e^{\frac{3JRT}{2P_T D}} + \frac{P_T}{3} \qquad (47)$$

Substitution of these equations into equation(28) yields an equation for the growth rate in terms of kinetic and diffusional parameters. The equation(28) can be solved for various values of the ratio of the partial pressures $P_A/P_{Bm}$ and the activation energies for adsorption of A and $B_m$. Since at this time there is little know data for AlN, to test our model we applied it to CdS and compared the result with the published papers.[30, 31, 32] The comparison showed a good match, which legitimized the model.

To model AlN, the value of diffusion coefficient used was 2.13 kg m/$s^3$. The partial pressure of components A and $B_m$ and the equilibrium constant $K_T$ are calculated from empirical data given by Dryburgh[33]. Simulated results are presented in the following chapter.

# CHAPTER IV: Results and Discussions

## 4.1 Temperature Dependence of Growth Rate:

Several growth experiments were performed to study the growth rate dependence on physical parameters such as temperature and pressure. The crucible structure described in figure 2.3 was employed in the experiments. A temperature gradient is imposed upon the crucible with the aid of the graphite support rod which acts to transport thermal energy to a heat sink at the top of the furnace. In the experimental results reported here, temperature gradients of $2.5^0$ - $4^0$ per mm were used. The growth experiments were performed in the temperature range of $1900^0$ - $2200^0$C with ambient nitrogen pressure of 200-500 torr. Optical pyrometer is used to measure the source temperature and an electric pressure gauge is used to calculate the total pressure of the system. The sublimation reaction of AlN can be represented as follows:

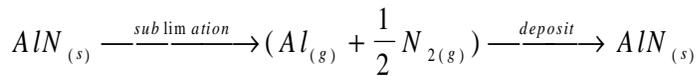

$$AlN_{(s)} \xrightarrow{sublimation} (Al_{(g)} + \frac{1}{2} N_{2(g)}) \xrightarrow{deposit} AlN_{(s)}$$

AlN was grown at different temperatures with constant pressure and temperature difference, and the thickness of the growth material was measured to calculate the growth rate. The temperature gradient is estimated from the temperature profile graph, described in section 2.4. The results thus obtained are presented in table 4.1 and table 4.2.

### Table 4.1: Growth rate as a function of Temperature

Total Pressure $P_T$ = 500 Torr

Temperature difference $\Delta T \approx 4.8^0 K$

| Temperature in $^0K$ | Growth Rate microns/h |
|---|---|
| 2430 | 37 |
| 2460 | 38 |
| 2480 | 41 |

### Table 4.2: Growth rate as a function of temperature

Total Pressure $P_T$ = 400 Torr

Temperature difference $\Delta T = 4.8^0 K$

| Temperature in $^0K$ | Growth Rate in micron/h |
|---|---|
| 2430 | 38 |
| 2460 | 47 |
| 2480 | 48 |

From the theoretical simulation, the relation between growth rate and temperature at constant pressure and temperature gradient can be estimated. Since the exact temperature gradient is difficult to calculate in the experiment, the program was developed to plot the behavior for multiple temperature gradients. In figure 4.1, the growth rate is plotted as a function of source temperatures. The figure comprises both the experimental results and the estimated results from our mathematical model. In figure 4.1, the growth rate is plotted as a function of source temperatures at constant pressure and temperature gradients. Figure 4.1 shows the relation at different temperature gradients. It comprises both the experimental results and the estimated results from our model. Growth rate strongly depends upon the source temperature. A close match is observed between the two results. It is very difficult to measure the exact temperature of the source. We used an optical pyrometer to measure the temperature of the source. This uncertainty in temperature measurement hampers our result slightly. Nevertheless, the graph indicates that the growth rate is almost the same as predicted by the model. We can infer that the model is quite good to predict the growth rate at different temperatures.

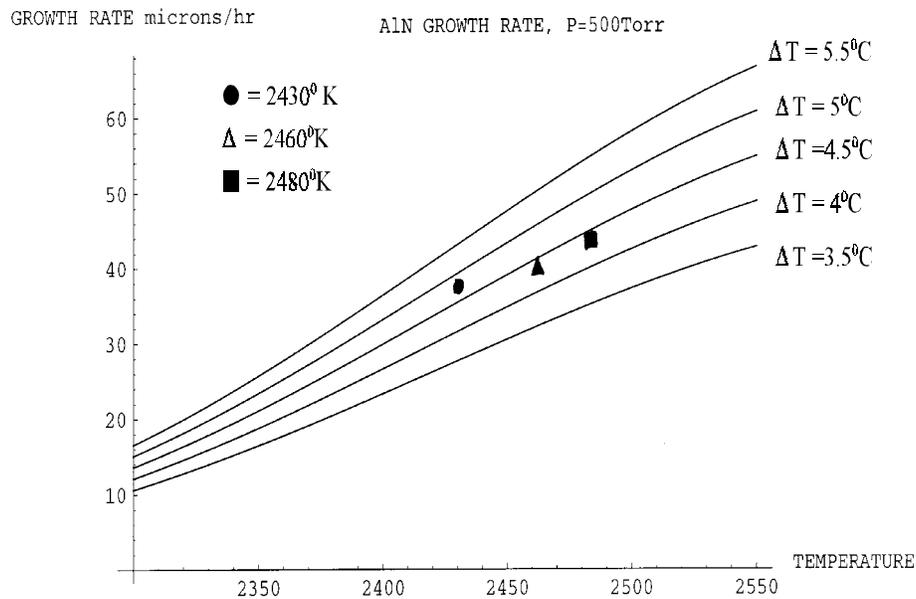

Figure 4.1  Growth rate as a function of temperature

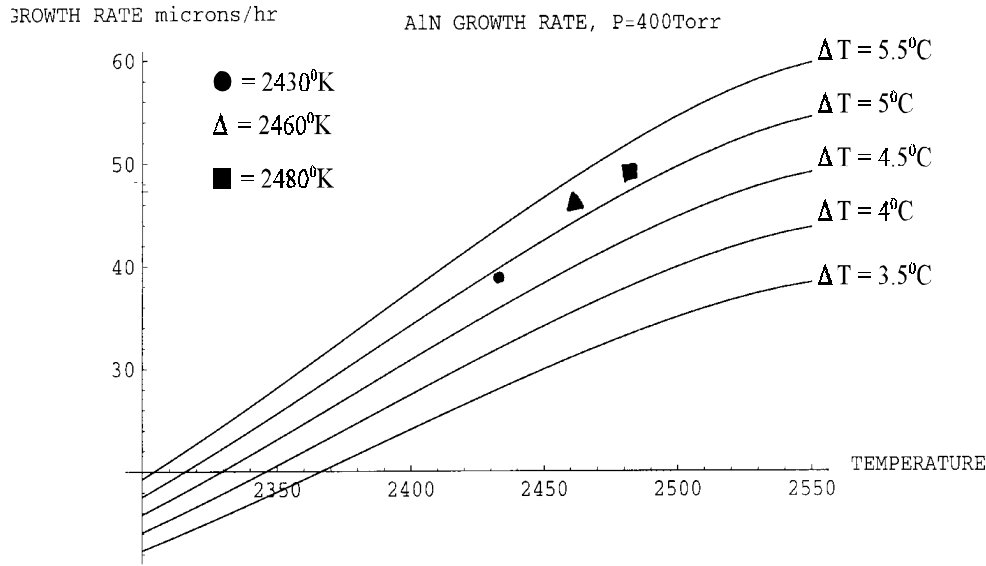

Figure 4.2 Growth rate as a function of temperature

## 4.2 Pressure Dependence of Growth Rate:

Growth pressure plays an important role in the determination of growth rate. The Al and $N_2$ vapor species in equilibrium with AlN can have different partial pressure ratios. It is very difficult to measure the exact partial pressure ratio of Al and $N_2$. Since the crystal is grown at a high nitrogen ambient pressure, the partial pressure of $N_2$ is close to the total pressure of the system. AlN was grown at different pressures at constant temperature and temperature gradient and the growth rate was calculated for the corresponding condition. The result obtained is presented in table 4.3 and table 4.4.

**Table 4.3: Growth rate as a function of pressure**

Temperature T = 2430°K

Temperature difference $\Delta T$ = 4.8°K

| Pressure in Torr | Growth Rate in micron/h |
|---|---|
| 200 | 32 |
| 300 | 36 |
| 400 | 38 |
| 500 | 37 |

**Table 4.4: Growth rate as a function of pressure**

Temperature T = 2460°K

Temperature difference $\Delta T$ = 4.8°

| Pressure in Torr | Growth Rate in micron/h |
|---|---|
| 200 | 33 |
| 300 | 39 |
| 400 | 47 |
| 500 | 41 |

For simplicity, we have calculated the value of equilibrium constant from Dryburg.[32] The growth rate as a function of pressure is plotted in figure 4.3. Both the experimental and the calculated results are presented. The results shows a fairly good match with the expected modeled behavior. There was an uncertainty in the temperature difference measurement $\Delta T$, as discussed in section 2.4. From the temperature profile it was found that $\Delta T$ tends to a lower value at higher temperature and smaller distance from the Z2 center. We have taken $4.8_{+0.45}^{-0.30}$ as temperature difference $\Delta T$, to plot the growth rate as a function of pressure as indicated by figure 4.2. There some uncertainty in the pressure measurement also. During the growth process, the total pressure of the system was maintained by opening and closing the mechanical pump of the system. This made it difficult to keep the pressure at a precise value, and the pressure varied 10 to 15 torr from the constant value. Despite these shortcomings, the growth rate matched quite well with the values predicted by the model. These results again indicate the validity of the model to estimate the growth rate.

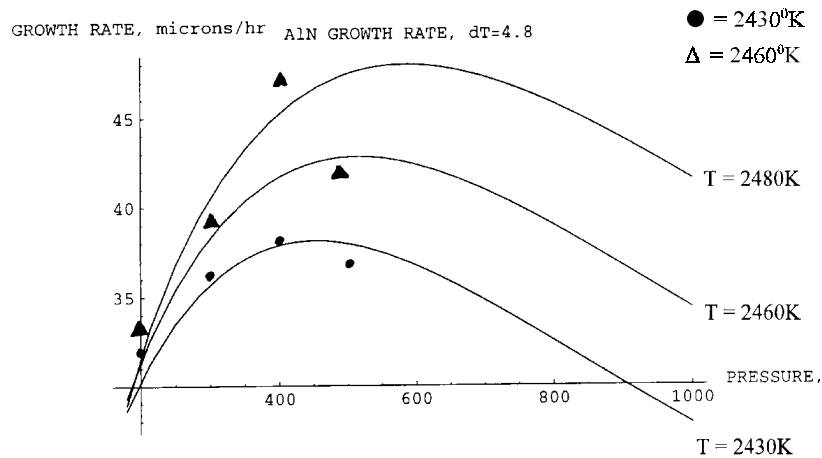

Figure 4.3a Growth rate as a function of pressure

A relation between the temperature of the source and the total pressure of the chamber at a maximum growth rate is plotted in figure 4.3. This relation helps to predict the appropriate growth pressure at a certain growth temperature or vice versa.

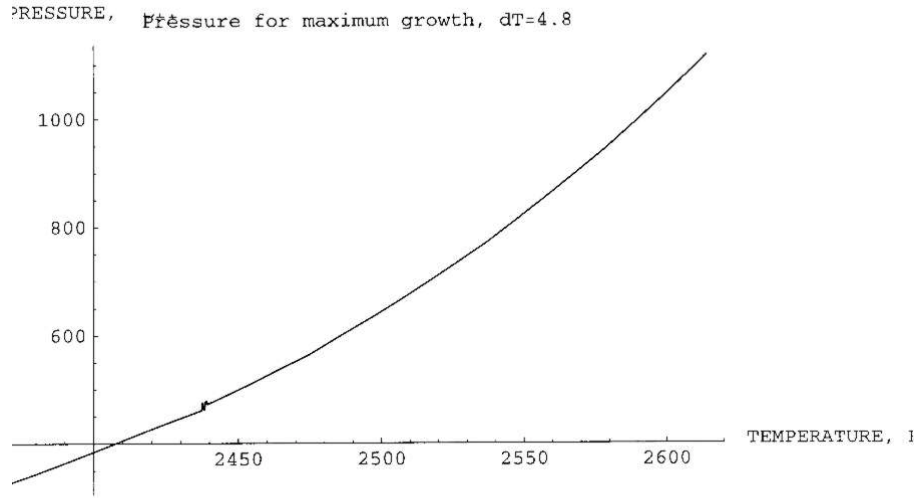

Figure 4.3b Pressure as a function of temperature at the maximum growth rate

Maximum growth rate as a function of temperature is shown in figure 4.4. This gives the experimental parameters to be maintained for optimal growth rates.

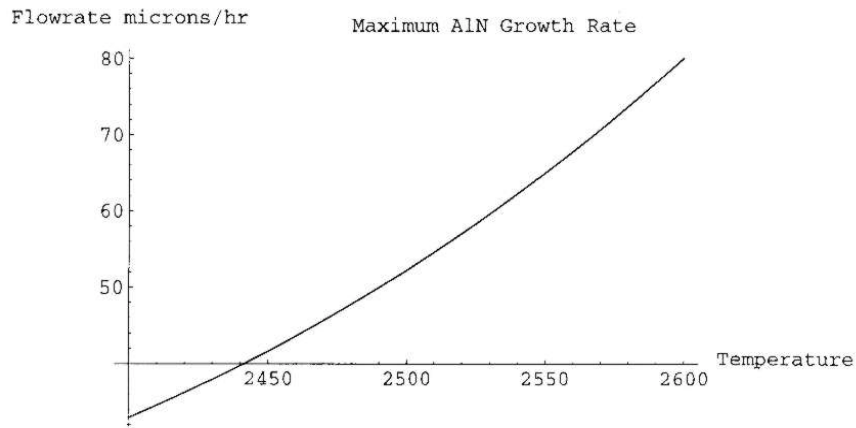

Figure 4.4 Maximum growth rate as a function of temperatures

# CHAPTER V: Further Experimental Results

The results pertaining to the surface studied using an optical microscope. The thickness of the growth layers varied form 10microns to 50microns. Crystal growth at source temperature in the range of 2157°C ~ 2182°C showed high degree of mono crystalline. A seed-source separation of 2-4 mm and system pressure of 500 torr was used in most of the growth experiments. As mentioned, bulk AlN is grown on both off-axis and on-axis 6HSiC substrates. We have observed step-flow growth and three-dimensional pyramid-like (hillocks growth) of AlN on off-axis and on-axis substrates respectively. Growth is full coverage of the substrate for both cases.

The Auger spectrum of the grown bulk AlN was found to contain fewer contaminants after the source and crucible modifications were made. These modifications were seen to significantly reduce the level of oxygen and crucible contamination. X-ray diffraction measurements were also performed on our bulk AlN sample, at A.F Ioffe Institute, St. Petersburg, Russia. X-ray diffraction measurements were conducted employing double crystal and triple crystal spectrometers. The X-ray rocking curve in $\omega$-$2\theta$ geometry was measured using a triple crystal spectrometer. Figure 5.1 shows double crystal rocking curve measured using $\omega$-$2\theta$ scanning geometry for (002) reflection of bulk AlN. The FWHM of this scanning geometry is 25.9 arcsec, which is a fairly good value indicating high degree of crystalline quality.

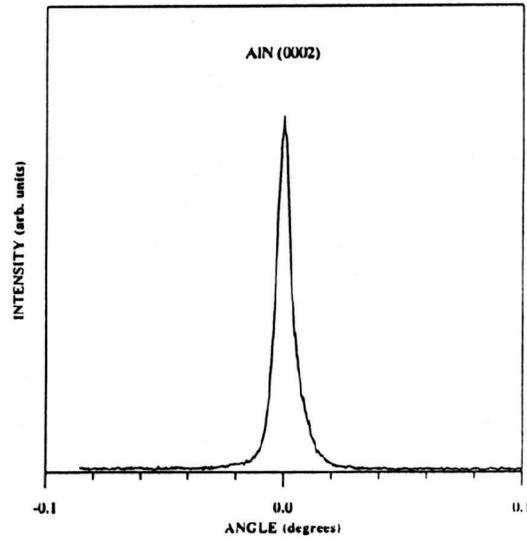

Figure 5.1 Rocking curve of AlN in $\omega - 2\theta$ scan

Figure 5.1 shows an asymmetry of the ω-2θ rocking curve. This asymmetry of the rocking curve (ω-2θ) is explained by a difference between lattice constants of AlN sub- layer near the interface and the top part of the layer. The low value of FWHM suggests that the growth material has fewer dislocations. The dislocation densities can also be measured directly by a transmission electron microscopy (TEM).

Transmission electron microscopy (TEM) was used to characterize our growth material (bulk AlN). A JEOL 4000 FX TEM operated at 300 KV is used to study the material. Cross-sectional TEM samples were prepared using tripod polishing and room temperature ion milling.

A TEM images of the bulk AlN grown on 3.5° off-axis 6HSIC is shown in figure 5.2.

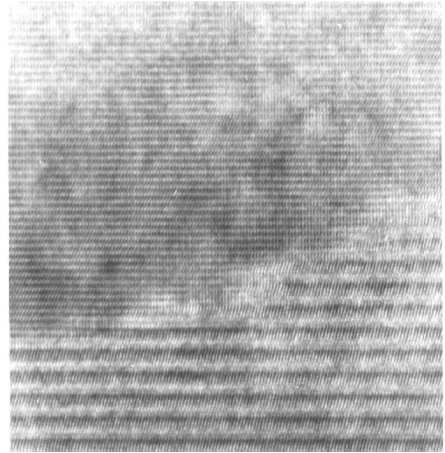

Fig. 5.2 High resolution lattice image of bulk AlN/6HSiC (off-axis) interface

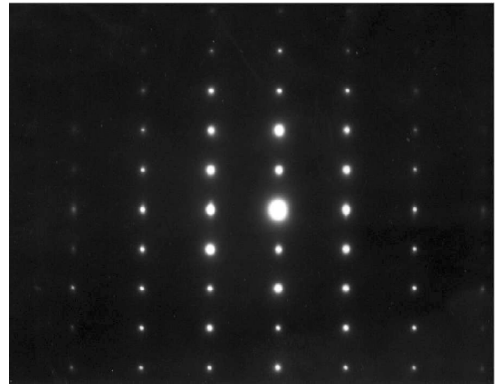

Fig. 5.3 Diffraction pattern of the bulk AlN

In spite of some bunching of the substrates' surface, the high-resolution image shows that the quality of the AlN/SiC interface is good, which suggests that the structural quality of the bulk AlN is quite good. Figure 5.3, shows the diffraction patterns of the bulk AlN, which are spot patterns. Since the spot pattern is an indicator of a single crystal, it can be infered that the growth material is single crystal. This means, that the AlN is well-aligned with the substrate.

Overall, we feel that our technique produces bulk AlN crystals of high quality, and that we have developed procedures which allow us to optimally grow and characterize such material with straightforward effort.

# Conclusion

In this work, an effort has been made to grow quality bulk AlN single crystal samples using physical vapor transport. A mathematical model was developed to optimize the growth process. Comparisons made between the experimental and the computational results have been used to determine the viability of the computational techniques.

We developed a mathematical model to simulate the system considering the different thermodynamic parameters involved in the system. In our model, both the diffusion and the activation process have been considered. It was noticed that the growth rate is limited by diffusion primarily at lower activation energy and at higher activation energy it is controlled primarily by the activation energy. The high activation energy indicates that AlN crystal growth from a vapor phase is primarily activation limited.

In spite of uncertainties of temperature measurements, the plots obtained indicate a close match between the experimental and estimated values. The form of the growth rate as a function of pressure displays a maxima for certain pressure ranges. It was noticed that the maximum growth lies between 300-500Torr. Despite the limitation of accurate pressure measurement, experimental result shows a fairly well matches with the estimated result.

For the experiments, we used AlN powder as the source material. Since nitrogen easily reacts with the atmospheric oxygen to form nitrous oxide, it was very difficult to avoid oxygen contamination of the source material. We have overcome this problem by reusing the same source material. It was observed after several growths that the AlN powder molded to formed an AlN chalk. We reused the AlN chalk as a source, with an idea that surface area of AlN chalk was less than AlN powder, giving the AlN in the chalk less likelihood to react with oxygen. A contamination free single crystal of AlN was obtained after we have made the change of source material from powder to molded AlN chalk. Others[17] have used AlN palate as a source material to grow AlN crystal, which is very expensive. We were able to grow better quality AlN single crystal using our source. To our knowledge no one has reported AlN chalk as a source to grow AlN crystal.

The second problem was the carbon contamination of the source material from the graphite crucible. We have used tungsten foil as a shield to protect the source from the crucible wall.

Another problem involved with cracking of the growth material, which is a common problem for nitride materials. It is believed that growth material cracks because of the mismatch of thermal expansion coefficients between the substrate and the growth material. To investigate the origin of the cracks, we used AlN crystal, grown by hydride vapor deposition, as a substrate. It was observed that the growth material had no cracks in it after the substrate material was changed. This suggests that the observed crack in the materials was indeed due to the mismatch of the thermal coefficients.

## Acknowledgments

The authors acknowledge Peizhen Zhou for her assistance in AES measurement; Dr. Vladimir, Yu Melnik, A. Nikolaev of the A.F. Ioffe Institute, in Russia for X-ray diffraction measurements; W.L. Sarney for her assistance in TEM measurement; Dr. Yuri Makarov for helpful discussions; Mr. James Griffin, Mr. Anthony Gomez, and Crowford Taylor, for their assistance in the maintenance of the sublimation reactor and valuable suggestions, and Afreen Afrose and Ahmed Arefeen Adil for help in typing the manuscript.

Finally, TH extends his gratitude to his parents and family members who made it possible for him to pursue higher studies in the United States of America.

# References

1) Nakamura, S., Mukai, T., and Senoh, M., *J. Appl. Phys.* 76. 8189 (1994)

2) Molnar, R.J., Singh, R., and Moustakas, T.D., *Appl. Phys. Lett.* 66, 268 (1995)

3) Khan, M.A., Chen, Q., Skogman, R.A., and Kuznia, J.N., *Appl. Phys. Lett.* 66, 2046 (1995)

4) Sverdlov, B.N., Martin, G.A., Markoc, H., *Appl. Phys. Lett.*, 67, 2063 (1995)

5) Thurmond, C.D., Logan, R.A., *J. Electrochem. Soc.* 119, 622 (1972)

6) Landolt and Bornstein, *Numerical Data and Fundamental Relationships in science and Technology*, Vol.17, Semiconductors, Springer, Berlin (1984)

7) Popovici, G., Morkoc, H., Mohammad, S.N., *Group III Nitride Semiconductor Compounds*, ed. Gil, B., Clarendon Press, Oxford (1998)

8) Hamilton, D. *J. Electrochem. Soc.* 105, 735, (1958)

9) Taylor, K. and Lenie, C. *J Electrochem. Soc.* 107, 1023, (1960)

10) Cher, T. Ing, D. and Noreika, A. *Sol. State Elect.* 10, 1023, (1967)

11) Kawabe, K. Tredgold, R. and Inuishi, Y. *Elect. Eng. Japan* 87, 62, (1967)

12) Drum, C. and Mitchell,J. *Appl. Phys. Lett.* 4, 164, (1964)

13) Wolf, J. Anorg, Z. *chem.* 87, 120, (1941)

14) Champbell, R.B. and Chang, H.C. *Solid State ultraviolet Devices for Fire Detection in Advanced Flight Vehicles, AD815895*, (1967) [*chem.* Abstr. 72, 94536r, (1970)]

15) Knippenberg, W.F. and Verspui, G. *U.S. Patent 3624*, 149, Jan. 11, 1972

16) Slack, G.A. and McNelly, T.F. *J. Crystal. Growth*, 42, 560, (1977)

17) Balkas, C.M. Sitar, Z. Zheleva, T. Bergman, L. Shmegrin, I.K. Muth, J.F. Kolbas, R. Nemanieh, R. and Daves, R.F. *Mat. Res. Symp. Proc.* Vol. 449, (1997)

18) Shields, V. Fekada, K. Johnson, P. Spencer, M. Aluko, M. Harris, G. Springer *Proceedings in physics*, vol 56 [Springer, Berling, Heidelberg, New York, 1992]

19) Lander, J.J. in *"Progress in Solid State chemistry,"* Vol. 2, p. 26, Pergamon Press, New York 1965

20) McGuire, G.E. *"Auger Electron Spectroscopy Reference Manual,"* Plenum Press New York (1979)

21) Sekine, T. Nagasawa, Y. Kudoh, M. Sakai, Y. Pardes, A.S. Geller, J.D. Mogami, A. and Kirata, K. *"Handbook of Auger Electron Spectroscopy",* Jeol., Tokyo (1982)

22) Bobb, L.C. Holloway, H. Maxwell, K.H. and Zimmerman, E. *J. Appl. Phys.* 37, 4687, (1966)

23) Guller, G.W. Corboy, J.F. and Smith, R.T. *J. Cryst. Growth* 31, 274, (1975)

24) Nobuo, Itoh. and Keiichi, Okamoto. *J. appl. Phys.* 63(5), 1486, (1988)

25) Feldman, L.C, Mayer, J.W., *Fundamentals of surface and thin film analysis*, 177, Elsevier Science Publishing Co.,Inc. 1986.

26) Caveney, R.J. *J. Crystal Growth* 7, 102, (1970)

27) Trapnell, B.M.W. *Chemisorption*, Butterworths scientific publication, London, 1955

28) Tompkins, F.C. *Chemisorption of Gasses on Metals*, Academic Press Inc. (London) LTD. 96, (1978)

29) Jeans, J. *"Kinetic Theory of Gases,"* The University Press, Cambridge, 1962

30) Faktor, M.M. Garrett, I. and Hechingbottom, R. *"Diffusional Limitations in a gas phase growth of crystals," J. Cryst. Growth*, 9, 3, (1971)

31) Faktor, M.M. Hechingbottom, R. and Garrett, I. *Growth of crystals from the fas phase. Part I. Diffusional Limitations and Interfacial Stability in crystal growth by dissociative sublimation," J. chem. Soc. (A).* 2657, (1970)

32) Faktor, M.M. and Garrett, I. *"Interplay of activation and diffusion in crystal growth from the vapour phase," J. cryst. growth* 9, 12, (1971)

33) Dryburg, *J. Cryst. Growth*, 125, 65-68, (1992)